
\input jnl
\input eqnorder
\input reforder

\def\3he{{$^3${\rm He}}}

\def\ie{{\it i.e.,\ }}

\def\slD{\raise.15ex\hbox{$/$}\kern-.53em\hbox{$D$}}
\def\dsl{\raise.15ex\hbox{$/$}\kern-.57em\hbox{$\Delta$}}
\def\slp{{\raise.15ex\hbox{$/$}\kern-.57em\hbox{$\partial$}}}
\def\nsl{\raise.15ex\hbox{$/$}\kern-.57em\hbox{$\nabla$}}
\def\sla{\raise.15ex\hbox{$/$}\kern-.57em\hbox{$\rightarrow$}}
\def\slla{\raise.15ex\hbox{$/$}\kern-.57em\hbox{$\lambda$}}
\def\slb{\raise.15ex\hbox{$/$}\kern-.57em\hbox{$b$}}
\def\lnp{\raise.15ex\hbox{$/$}\kern-.57em\hbox{$p$}}
\def\lnk{\raise.15ex\hbox{$/$}\kern-.57em\hbox{$k$}}
\def\lnK{\raise.15ex\hbox{$/$}\kern-.57em\hbox{$K$}}
\def\lnq{\raise.15ex\hbox{$/$}\kern-.57em\hbox{$q$}}

\def\chib{{\overline\chi}}
\def\psib{{\overline\psi}}


\def\pmb#1{\setbox0=\hbox{$#1$}%
\kern-.025em\copy0\kern-\wd0
\kern.05em\copy0\kern-\wd0
\kern-.025em\raise.0433em\box0 }

\def\q2{{Q^2}}
\def\gtwid{\raise.3ex\hbox{$>$\kern-.75em\lower1ex\hbox{$\sim$}}}
\def\ltwid{\raise.3ex\hbox{$<$\kern-.75em\lower1ex\hbox{$\sim$}}}
\def\12{{1\over2}}
\def\part{\partial}

\def\low#1{\lower.5ex\hbox{${}_#1$}}

\def\psl{\raise.15ex\hbox{$/$}\kern-.57em\hbox{$\partial$}}
\def\partt{\raise.15ex\hbox{$\widetilde$}{\kern-.37em\hbox{$\partial$}}}

\def\topppageno1{\global\footline={\hfil}\global\headline
={\ifnum\pageno<\firstpageno{\hfil}\else{\hss\twelverm --\ \folio
\ --\hss}\fi}}

\def\toppageno2{\global\footline={\hfil}\global\headline
={\ifnum\pageno<\firstpageno{\hfil}\else{\rightline{\hfill\hfill
\twelverm \ \folio
\ \hss}}\fi}}

\def\prl#1{Phys.\ Rev.\ Lett.\ {\bf #1}}
\def\prd#1{Phys.\ Rev.\ {\bf D#1}}
\def\plb#1{Phys.\ Lett.\ {\bf #1B}}
\def\npb#1{Nucl.\ Phys.\ {\bf B#1}}

\def\ie{{\it i.e.},\ }

\def\ref#1{${}^{#1}$}
\def\nsection#1 #2{\leftline{\rlap{#1}\indent\relax #2}}

\def\prl#1{Phys.\ Rev.\ Lett.\ {\bf #1}}
\def\prd#1{Phys.\ Rev.\ {\bf D#1}}
\def\plb#1{Phys.\ Lett.\ {\bf #1B}}
\def\npb#1{Nucl.\ Phys.\ {\bf B#1}}

{\parindent=0pt
{Sept. 1994}\hfill{Wash. U. HEP/94-61}
\rightline{WIS--94/37--PH}
\rightline{hep-lat/9409013}
}
\title Domain wall fermions in a waveguide: the phase diagram at large
       Yukawa coupling
\author Maarten F.L. Golterman$^{a,*}$ and Yigal Shamir$^{b,**}$
\affil
$^a$Department of Physics, Washington University
St.~Louis,~MO~63130,~USA
\null
$^b$Department of Physics, Weizmann Institute of Science
Rehovot~76100,~Israel

\abstract
{In this paper we return to a model with domain wall fermions in
a waveguide.  This model contains a Yukawa coupling $y$ which is needed
for gauge invariance.  A previous paper left the analysis for large
values of this coupling incomplete.  We fill the gap by developing
a systematic expansion suitable for large $y$, and using this, we
gain an analytic understanding of the phase diagram and fermion spectrum.
We find that in a sense all the species doublers come back for large
$y$. The conclusion that no lattice chiral gauge theory can be
obtained from this approach therefore remains valid.}
\vfill
\item{$^*$} e-mail: maarten@aapje.wustl.edu
\item{$^{**}$}{address after Oct. 1st: School of Physics and Astronomy,
Tel-Aviv Univ., Ramat Aviv, 69978 Israel; e-mail:
ftshamir@wicc.weizmann.ac.il}
\endtopmatter

\def\I{{\it I}}
\def\cf{{\it cf.\ }}
\def\ie{{\it i.e.\ }}
\def\eq#1{{eq. \(#1)}}

\def\psib{{\overline{\psi}}}
\def\Psib{{\overline{\Psi}}}
\def\chib{{\overline{\chi}}}
\def\del{\partial}
\def\dsl{\del\!\!\!/\,}
\def\Dsl{D\!\!\!\!/\,\,}
\def\psl{p\!\!\!/\,}
\def\ssl{s\!\!\!/\,}
\def\half{{{1\over 2}}}
\def\M{{\cal M}}
\def\square#1#2{{\vcenter{\vbox{\hrule height.#2pt
    \hbox{\vrule width.#2pt height#1pt \kern#1pt \vrule width.#2pt}
    \hrule height.#2pt}}}}
\def\Box{{\mathchoice{\sl\square63}{\sl\square63}\square{2.1}3\square{1.5}3}\,}

\subhead{\bf 1. Introduction}

A recent proposal to obtain $d=2n$ dimensional lattice chiral fermions
from a domain wall in $d+1$ dimensions [\cite{kaplan}] has generated a lot
of interest. The basic observation is that if one introduces a domain
wall in an odd dimensional theory of a Dirac fermion by introducing a
mass parameter which changes sign across a hyperplane of codimension
one (the ``domain wall"),
massless states occur that are bound to this domain wall. These
states therefore can be interpreted as to live in one dimension less, if
one takes the size of the mass parameter to be of the order of the cutoff
(the spread of the bound state wavefunctions is of order of the
inverse mass in the direction orthogonal to the hyperplane).
Moreover, these states carry a definite chirality, which depends on the
overall sign of the mass parameter.

Kaplan realized that the fact that the odd dimensional Dirac theory is
entirely vectorlike in nature, without any chiral symmetry appearing at
the level of the action, can be used to implement this idea on the
lattice.  Because the action is vectorlike, a Wilson mass term can be
introduced without breaking any symmetries of the
action.  This Wilson term eliminates the well known species doublers
that are generated on the lattice.  If we denote the coordinates of the
$d$ dimensional space by $x$, and that of the extra $d+1$-st coordinate
by $s$, a local (\ie single site) mass term is added which is
positive for $s>0$ and negative for $s<0$.  Kaplan showed
that in this way
the above described phenomenon carries over to the lattice, and
that a chiral fermion emerges bound to the domain wall defect located at
$s=0$.

In a sense the theory of a massless chiral fermion constructed in this
way is not entirely free of species doublers. If we consider the theory
in a finite but large volume with (anti)periodic boundary conditions, an
antidomain wall defect has to occur somewhere, and a chiral fermion with
chirality opposite to that of the massless fermion bound to the domain
wall occurs at the antidomain wall. In a finite volume the wavefunctions
of these two modes have a small overlap (approximately
exponentially suppressed as
$exp(-mL)$, where $m$ is the absolute value of the mass parameter, and $L$
is the distance between the domain wall and the antidomain wall), and a
tiny Dirac mass couples the two modes, which however become massless and
decoupled in the infinite volume limit. This phenomenon was shown to be
quite general (\ie independent of the precise choice of boundary
conditions) by Narayanan and Neuberger [\cite{narneu}]. An alternate,
equivalent
construction employing free boundary conditions on a odd dimensional
half space was proposed and discussed by one of us [\cite{shamir}].

This theory was first studied in the presence of background gauge fields
in order to assess the role of the anomaly.  In the original
paper [\cite{kaplan}], it was argued that a Goldstone--Wilczek current
flowing off the domain wall should exist to reconcile the apparent
conflict between the anomalous structure of a single even dimensional
chiral fermion and the exact gauge invariance of the odd dimensional
theory, much like the mechanism discussed by Callan and Harvey in
the continuum case [\cite{calhar}].  This was indeed confirmed in a
series of papers [\cite{jansen,goljankap,crehor}].  In particular, it was
shown in ref.  [\cite{goljankap}] that a Goldstone--Wilczek current is
generated in the region where the combined mass matrix (consisting of the
local domain wall mass and the Wilson mass terms) is not positive
definite, carrying anomalous charge between the domain wall and the
antidomain wall in a way consistent with the chiral zeromode spectrum at
those defects [\cite{janschm}].

The situation becomes more complicated when one considers dynamical
gauge fields. Obviously, one would like to introduce gauge fields
that do not couple both
to the zeromodes living at the antidomain
wall and also to those at the domain wall, since this would render the
resulting gauge theory vectorlike. The task is therefore to couple gauge
fields in such a way that they do not couple to the fermion modes living
at the antidomain wall. A concrete implementation has been proposed in
which the gauge fields are confined to a ``waveguide" around the domain
wall, excluding a region around the antidomain wall [\cite{kaplan2,
paper1}]. Within this waveguide, the gauge fields are taken to be $d$
dimensional, \ie the gauge fields are independent of the direction
orthogonal to the waveguide (the $s$ coordinate), and all link variables
in this extra direction are equal to one [\cite{narneu,kaplan2}]. (For a
different approach, where the extent of space in the $s$-direction is
kept strictly infinite, see refs.  [\cite{narneu2}]. See also ref.
 [\cite{shamir2}].)

By introducing a waveguide, new defects are necessarily introduced into
the theory, consisting of the boundaries of the waveguide. These are $d$
dimensional hyperplanes situated somewhere ``halfway" between the domain
wall and the antidomain wall. With the specific choice of gauge fields
as described above, gauge invariance is explicitly broken at these
boundaries. This is remedied by introducing a $d$ dimensional scalar
field that lives only at the two boundaries, restoring gauge invariance
in a way similar to the way fermion mass terms are made gauge invariant
in the electroweak standard model. The Yukawa coupling between this
scalar field and the fermion field is a free parameter of the model. It
is important to point out that within the context of the waveguide
model, such scalar fields are unavoidable. They are essentially the
gauge degrees of freedom, which, due to the explicit breaking of gauge
invariance, couple to the fermions. Moreover, their dynamics is
necessarily nonperturbative, since their fluctuations are not controlled
by a small parameter (such as the gauge coupling, which only controls
the transverse degrees of freedom) [\cite{janzak,janseillac}].

The key question is now of course whether these new defects will lead to
new fermionic zeromodes. Any such new zeromodes bound to the inside of a
waveguide boundary will couple to the gauge field, and presumably
destroy the chiral nature of the theory. In fact, by confining the gauge
field to the waveguide, no Goldstone--Wilczek current can reach the
antidomain wall anymore, and one may legitimately worry whether new
chiral zeromodes are needed at the waveguide boundaries in order to
absorb the anomalous charge carried by the Goldstone--Wilczek current
between the domain wall and the boundary in the region with nonpositive
definite fermion mass matrix. However, it was argued in ref.
 [\cite{paper1}] (which from now on we will refer to as \I), that this is
not necessarily the case, and that the existence or nonexistence of new
fermionic zeromodes at the waveguide boundaries is a dynamical question.

In order to answer this question one needs to map out the phase diagram
and spectrum
of the theory.  This was attempted in \I, in which the waveguide model
without gauge fields, but with the scalar field, was studied analytically
and numerically as a function of the Yukawa coupling $y$ and the scalar
hopping parameter $\kappa$ (the scalar field was taken to be radially
frozen).  It was found that always new zeromodes appear on the waveguide
boundary in the region of nonpositive definite mass matrix, providing
``mirror" modes to the chiral modes living on the domain wall in a
one-to-one correspondence.

However, as described in \I, the numerical simulations were notoriously
difficult for large values of the Yukawa coupling, due to a bad signal
to noise ratio for the fermion propagators in that region, and it was
hard to arrive at any definite conclusion for large $y$.  In the present
paper we return to the question of the fermionic zeromode spectrum at
the waveguide boundaries for large Yukawa coupling.  It is well known
that in general new phases occur for large values of the Yukawa coupling
of lattice scalar-fermion models.  Those phases can be distinguished
from those at  small values of the Yukawa coupling by the fermionic
zeromode spectrum.  In particular, it is in principle possible that no
zeromodes occur at all at large $y$.  This would be an important
ingredient for constructing a genuinely chiral lattice gauge theory
from domain wall fermions.

Here, we develop
an expansion suitable for investigating the large $y$ region
of the phase diagram. Again, we set the gauge field link variables equal
to one, but keep the scalar field at the waveguide boundary. This allows
us to find the zeromode spectrum for large $y$. We do find a new
symmetric phase which is not connected to the symmetric phase at small
values of $y$.  However, the
zeromode spectrum in fact turns out to be very rich, and quite similar
to that of the Smit-Swift [\cite{janzak,jannpb,swift}] model for {\it
weak} Yukawa coupling: all the doublers of the zeromode at the domain
wall reappear at the waveguide boundaries!

In section 2 we define the waveguide model restricting ourselves to the
case $d=4$, and recast it in a way amenable to an expansion in $1/y$. In
the next section we find the zeromode spectrum for $y=\infty$, and in
section 4 we estimate the location of the phase transition between the
symmetric and broken phases in the $\kappa$--$y$ plane for large $y$. We
then briefly consider the zeromode spectrum at large but finite $y$.
In the last section we present our
conclusions.

\subhead{\bf 2. The model}

The waveguide implementation of the domain wall fermion model is defined
by the action
$$\eqalignno{
S_{\rm domain wall} = & \sum_{s\in WG} \Psib_s(\Dsl(U) -W(U) + m_s)\Psi_s
 + \sum_{s\not\in WG}  \Psib_s(\dsl-w+ m_s)\Psi_s \cr
  & + \sum_{s\not=s_0'-1,s_0}[\Psib_sP_R\Psi_{s+1}
                         + \Psib_{s+1}P_L\Psi_s ]-\sum_s \Psib_s\Psi_s
&(faction)\cr
 & +   y(\Psib_{s_0'-1}V^\dagger P_R\Psi_{s_0'}  +
\Psib_{s_0'}VP_L\Psi_{s_0'-1})
  +   y(\Psib_{s_0}VP_R\Psi_{s_0+1} + \Psib_{s_0+1}V^\dagger P_L\Psi_{s_0}).
}$$
In this equation we have suppressed the four dimensional coordinates,
and only indicated the fifth coordinate explicitly.
$\partial_\mu$ is the four dimensional
symmetric nearest neighbor difference operator, and $w$
denotes the four dimensional Wilson mass term. $D_\mu(U)$ and $W(U)$ are
gauge covariant versions of these. $m_s$ is the domain wall mass, which
is taken to be positive for $s>0$ and negative for $s<0$. (We can take
$m(0)=0$ for instance.)  $V$ is a
radially frozen scalar field which takes its values in the gauge group
$G$. The waveguide $WG$ consists of all lattice points with $s_0'\le
s\le s_0$, and the domain wall at $s=0$ is inside the waveguide (\ie
$s_0'<0$ and $s_0>0$). With periodic boundary conditions, there will be
an antidomain wall outside the waveguide, where $m_s$ again changes
sign. We have chosen the Wilson parameter $r=-1$, and the projection
operators $P_{R(L)}$ are $P_{R(L)}=\half(1+(-)\gamma_5)$. For full details,
see \I. (Here we have chosen some conventions different from those in \I,
where for instance we chose $r=1$.)
This model contains a righthanded massless fermion bound to the domain wall
and a lefthanded massless fermion bound to the antidomain wall,
irrespective of the value of $y$ (\cf \I).

In the present paper we are interested in the fermionic zeromode
spectrum at the waveguide boundaries, and
we will restrict ourselves to a simpler model in
which only the waveguide boundary between $s=s_0$ and $s=s_0+1$
is present, with no other
defects.  We imagine all other defects to be very far away.
Also, as in \I, we will choose $U=1$, \ie we will turn off the gauge
fields.
Including a hopping term for the
scalar field, the action for this situation is
$$S=S_F+S_B,\eqno(fulls)$$
with
$$\eqalignno{
S_F=&\sum_{x,y,s}\left(\Psib_{x,s}\dsl_{xy}\Psi_{y,s}
                     +\half\Psib_{x,s}\Box_{xy}\Psi_{y,s}
                     +m\Psib_{x,s}\Psi_{x,s}\right)\cr
    &+\sum_{x,s\ne s_0}\left(\Psib_{x,s}P_R\Psi_{x,s+1}
                      +\Psib_{x,s+1}P_L\Psi_{x,s}\right)
    -\sum_{x,s}\Psib_{x,s}\Psi_{x,s}\cr
    &+y\sum_x\left(\Psib_{x,s_0}VP_R\Psi_{x,s_0+1}
                   +\Psib_{x,s_0+1}V^\dagger P_L\Psi_{x,s_0}\right)
&(action)
}$$
and
$$S_B=-\kappa\sum_{x,\mu}{\rm tr}(V_xV^\dagger_{x+\mu}+h.c.),\eqno(sboson)$$
where
$$\eqalignno{
\dsl_{xy}&=\half\sum_\mu\gamma_\mu(\delta_{x+\mu,y}-\delta_{x-\mu,y}),
&(dslash)\cr
\Box_{xy}&=\sum_\mu(\delta_{x+\mu,y}+\delta_{x-\mu,y}-2\delta_{x,y}).
&(box)
}$$
We will always choose $0<m<1$.

This action is invariant under the transformations
$$\eqalignno{
\Psi_{x,s}&\to g\Psi_{x,s},\ \ s\le s_0,\cr
\Psi_{x,s}&\to h\Psi_{x,s},\ \ s\ge s_0+1,&(symmetry)\cr
V&\to gVh^\dagger,
}$$
where $g$ and $h$ are elements of $G$.

For $y$ small and in the symmetric phase ({\it i.e.} $\kappa<\kappa_c$)
one massless lefthanded fermion exists just inside the
boundary, and a righthanded one just outside, with wavefunctions which
fall off exponentially away from the boundary.  In the broken phase,
these two chiral modes pair up into a massive Dirac fermion with mass
proportional to $yv$ with $v$ the vacuum expectation value of the
field $V$. Combined with the zeromodes at the domain wall and the
antidomain wall, and turning on the gauge fields, this model results
in a vectorlike gauge theory for small $y$ (\cf \I).

We would now like to investigate the situation for large $y$.  An
expansion in $1/y$ can be developed by a slight rewriting of the action,
\eq{action}.
First, we will relabel the fermion variables
$$\eqalignno{
\chi_s&=\Psi_s,\ \ s<s_0,\cr
\chi_{Rs_0}&=\Psi_{Rs_0},\ \ \chi_{Ls_0}=\Psi_{Ls_0+1},&(redef)\cr
\chi_s&=\Psi_{s+1},\ \ s>s_0,\cr
\psi_R&=\Psi_{R,s_0+1},\ \ \psi_L=\Psi_{L,s_0},
}$$
where $\Psi_{R,L}=P_{R,L}\Psi$, {\it
etc}.  Note that $\psi$ is a four dimensional fermion field, \ie
independent of $s$.
Rescaling
$$
\psi\to{1\over{\sqrt{y}}}\psi,
\eqno(rescale)$$
the fermionic action becomes
$$\eqalignno{
S_F=&\sum_s\left(\chib_s\dsl\chi_s+\chib_sa(s)(\Box+m-1)\chi_s\right)
&(newaction)\cr
&+\sum_s\left(\chib_{Ls}\chi_{Rs+1}+\chib_{Rs+1}\chi_{Ls}\right)
+\psib_LV\psi_R+\psib_RV^\dagger\psi_L\cr
&+\sqrt{\alpha}\Bigl(\psib(\Box+m-1)\chi_{s_0}+
\chib_{s_0}(\Box+m-1)\psi\Bigr)
+\alpha\psib\dsl\psi,
}$$
with
$$\alpha={1\over y}\eqno(alpha)$$
and
$$a(s)=1-\delta_{s,s_0}.\eqno(as)$$
In \eq{newaction} we again suppressed all $x$ dependence.

\subhead{\bf 3. The fermion spectrum on the boundary for $y\to\infty$}

For $y\to\infty$, $\psi$ decouples, and the Dirac equation for
$\chi_s(p)=\sum_xe^{-ipx}\chi_{x,s}$
is
$$i\ssl(p)\chi_s(p)+a(s)(m-1-F(p))\chi_s(p)+P_R\chi_{s+1}(p)
+P_L\chi_{s-1}(p)=0,
\eqno(dirac)$$
with $\ssl(p)=\sum_\mu\gamma_\mu\sin{p_\mu}$ and
$F(p)=\sum_\mu(1-\cos{p_\mu})$.
We wish to look for solutions of this equation
(in Minkowski space) which also satisfy the
four dimensional massless Dirac equation, and which have a definite
chirality.  For such solutions
$$\eqalignno{
&\chi_{Rs+1}(p)=a(s)(1-m+F(p))\chi_{Rs}(p),&(solution)\cr
&\chi_{Ls-1}(p)=a(s)(1-m+F(p))\chi_{Ls}(p).
}$$
These equations have normalizable solutions bound to the waveguide boundary
due to the fact that $a(s_0)=0$.  The first equation has solutions
inside the waveguide ({\it i.e.} $\chi_{Rs}=0$ for $s>s_0$) for
$F(p)>m$.  Likewise, the equation for $\chi_L$ has
solutions outside ({\it i.e.} $\chi_{Ls}=0$ for $s<s_0$) for $F(p)>m$.

This leads to a rich spectrum of chiral zeromodes at the waveguide
boundary.  Let
$$\eqalignno{
\pi_A&\in {\cal E}\cup{\cal O},&(pisubA)\cr
{\cal E}&=\{(0,0,0,0),(\pi,\pi,0,0),\dots,(\pi,\pi,\pi,\pi)\}\cr
{\cal O}&=\{(\pi,0,0,0),(0,\pi,0,0),\dots,(0,\pi,\pi,\pi)\}.
}$$
The four dimensional Dirac equation has relativistic (continuum)
solutions for $p=\pi_A+{\tilde p}$ with ${\tilde p}\to 0$.  The
chirality gets flipped if $\pi_A\in {\cal O}$, {\it i.e.}
$\chi_R(p)$ is lefthanded when $p\approx\pi_A\in {\cal O}$
[\cite{jan}].  The
condition $F(p)>m$ excludes the modes around $p=0$, and results in a
spectrum of $7$ righthanded and $8$ lefthanded zeromodes inside the
waveguide boundary, and the mirror reflection of this outside the
boundary.  With the single righthanded zeromode on the domain
wall and the lefthanded one on the antidomain wall the zeromode
spectrum is nonchiral both inside and outside the waveguide.  The fact
that the spectrum outside is a mirror copy of the spectrum inside is in
accordance with a symmetry of the action:
$$\eqalignno{
\chi_{x,s}&\to\gamma_4\chi_{Px,2s_0-s},\cr
\psi_x&\to\gamma_4\psi_{Px},&(parity)\cr
V_x&\to V^\dagger_{Px},
}$$
where $Px=(-x_1,-x_2,-x_3,x_4)$.

The other boundary of the waveguide can be treated in exactly the same
way, with now however $m$ replaced by $-m$, since this boundary is located
on the other side of the domain wall.  The condition for the existence
of zeromodes becomes $F(p)>-m$ on both sides of this boundary, which
is satisfied for all $p$.  It follows that there are $8$ righthanded and
$8$ lefthanded zeromodes both inside and outside this boundary for $y\to
\infty$.

We will later need the $\chi$-fermion propagator for $y\to\infty$, which
can be found as the inverse of the Dirac operator
$$\left([i\ssl(p)+a(s)(m-1-F(p))]\delta_{s,s''}+P_R\delta_{s+1,s''}
+P_L\delta_{s-1,s''}\right)S^{(0)}_{s'',s'}(p)=\delta_{s,s'}.\eqno(diracop)$$
We find, for $s'=s_0$
$$\eqalignno{
S^{(0)}_{s,s_0}(p)&={{-i\ssl(p)+z(p)-b(p)}\over
{s^2(p)+\M^2(p)}}z^{s-s_0}(p)P_L,\ \ s<s_0,\cr
S^{(0)}_{s,s_0}(p)&={{-i\ssl(p)+z(p)-b(p)}\over
{s^2(p)+\M^2(p)}}z^{s_0-s}(p)P_R,\ \ s>s_0,&(propagator)\cr
S^{(0)}_{s_0,s_0}(p)&={{-i\ssl(p)}\over{s^2(p)+\M^2(p)}},
}$$
where
$$\eqalignno{
b(p)&=1-m+F(p)>0,\ \ \forall p,&(definitions)\cr
z(p)&={{1+s^2(p)+b^2(p)+\sqrt{(1+s^2(p)+b^2(p))^2-4b^2(p)}}
\over{2b(p)}}>1,\ \ \forall p,\cr
\M^2(p)&=1-b(p)/z(p)\ge 0,\ \ \forall p.
}$$
Let us check that this propagator gives rise to the same
zeromode spectrum as we found from the Dirac equation.  For $p=\pi_A$
we have $b(p)=2n+1-m$, with $n$ equal to the number of components of
$\pi_A$ that are equal to $\pi$.  This leads to $z=1/(1-m)$ for $n=0$
and $z=2n+1-m$ for $n\in\{1,2,3,4\}$, and hence the propagator has a
pole at $p=\pi_A$ when $\pi_A\ne 0$, but not for $p=0$.

For the other waveguide boundary $b(p)=1+m+F(p)$, and in this case
$z=2n+1+m$ for all $n$, so that the propagator has massless poles at all
$\pi_A$, in accordance with what we found earlier.

\subhead{\bf 4. The phase diagram for large $y$}

In the limit $y\to\infty$ the fields $\chi$, $\psi$ and $V$ decouple
from each other, and the
phase diagram of the model is dictated by $S_B$, in particular by the
bosonic hopping parameter $\kappa$.  For $G=U(1)$ or $SU(2)$ there will
be a second order phase transition at $\kappa=\kappa_c$.  (The fermionic
part of the action depends on $V$ even for $y\to\infty$, but the fermion
determinant is independent of $V$ in this limit.)  The theory consists
of a scalar field in the broken or symmetric phase, and free fermionic
modes, described by the field $\chi$.

We would now like to see how this changes if we turn on the coupling
$\alpha$.  The effective action for $V$ can be obtained by integrating
out the fields $\psi$ and $\chi$ in an expansion in $\alpha$.  Let us
first integrate out $\chi$, which can be done exactly, resulting in
an effective action for $\psi$:
$$\eqalignno{
S_{\psi}=&\sum_x\left(\psib_{Lx}V_x\psi_{Rx}+\psib_{Rx}V^\dagger_x
\psi_{Lx}\right)&(spsi)\cr
&+\alpha\sum_{xy}\left(\psib_x\dsl_{xy}\psi_y
-[\psib(\Box+m-1)]_xS^{(0)}_{(x,s_0),(y,s_0)}[(\Box+m-1)\psi]_y\right),
}$$
with
$$S^{(0)}_{(x,s_0),(y,s_0)}=\int_pe^{ip(x-y)}S^{(0)}_{s_0,s_0}(p),
\eqno(momprop)$$
where
$$\int_p=\int{{d^4p}\over{(2\pi)^4}}.\eqno(momint)$$
To order $\alpha^2$ the effective action for $V$ obtained by integrating
out all fermion fields then becomes
$$\eqalignno{
S_{eff}(V)&\equiv S_B(V)&(seff)\cr
&\phantom{=}-\log\det\left[\delta_{x,y}+\alpha(V^\dagger_xP_R+V_yP_L)
\left(\dsl_{x,y}
-[(\Box+m-1)S^{(0)}_{s_0,s_0}(\Box+m-1)]_{x,y}\right)\right]\cr
&=-\left(\kappa+\half\alpha^2\left(1-\int_p{{s^2(p)b^2(p)}\over
{s^2(p)+\M^2(p)}}\right)\right)\sum_{x,\mu}{\rm
tr}(V_xV^\dagger_{x+\mu}+h.c.)\cr
&\phantom{=}-2\alpha^2\sum_{xy}{\rm tr}(V^\dagger_xV_y)\sum_\mu\int_{pq}
e^{i(p-q)(x-y)}s_\mu(p)s_\mu(q){{b^2(p)}\over
{s^2(p)+\M^2(p)}}{{b^2(q)}\over{s^2(q)+\M^2(q)}}\cr
&\phantom{=}+O(\alpha^4)
}$$
(recall $s_\mu(p)=\sin{p_\mu}$).

There will be a similar contribution from the waveguide boundary on the
other side of the domain wall, with the only difference that $m$ be
replaced everywhere by $-m$.

This effective action contains long range self-interactions for the field
$V$, due to the fact that the integrand of the last term has massless
poles at $p=\pi_A\ne 0$ (and for all $\pi_A$ for the terms with $m\to
-m$).
It is well known that a meanfield approach gives the correct qualitative
phase diagram for $\alpha=0$.  While this is in general not true for
more complicated interactions, we expect it to hold for $\alpha$ small
enough.  If we apply meanfield, we obtain an estimate for the location
of the critical line in the $\kappa$--$\alpha$ plane (valid for small
$\alpha$):
$$\kappa+\half\alpha^2\left[I(m)+I(-m)\right]
+O(\alpha^4)=\kappa_c,\eqno(critline)$$
with
$$I(m)=1-\int_p{{s^2(p)b^2(p)}\over
{s^2(p)+\M^2(p)}}+\half\int_p{{s^2(p)b^4(p)}
\over{[s^2(p)+\M^2(p)]^2}},\eqno(integral)$$
which is a function of $m$ through $b$ and $\M$.  As an example,
the value of $I$
is $187.4(3)$ for $m=0.5$ and $371.3(6)$ for $m=-0.5$.  The large values
of $I(m)$ reflect the enhanced coupling of the fermions to the $V$ field
(\cf the factors $b(p)$).

\subhead{\bf 5. Fermion spectrum for large but finite $y$}

We can integrate out the field $\psi$ in an expansion in $\alpha=1/y$
in order to obtain an effective action for $\chi$ and $V$:
$$\eqalignno{
S_\chi=&\sum_s\left(\chib_s\dsl\chi_s+
\chib_sa(s)(\Box+m-1)\chi_s\right)
+\sum_s\left(\chib_{s}P_R\chi_{s+1}+\chib_{s+1}P_L\chi_{s}\right)
&(schi)\cr
&-\alpha\sum\chib_{s_0}(\Box+m-1)\sum_{n=0}^\infty(-\alpha)^n
[(V^\dagger P_R+VP_L)\dsl]^n(V^\dagger P_R+VP_L)(\Box+m-1)\chi_{s_0}\cr
=&\sum_{s,s'}\chib_sS^{(0)-1}_{s,s'}\chi_{s'}
-\alpha\sum_x[\chib_{s_0}(\Box+m-1)]_x(V^\dagger_xP_R+V_xP_L)
[(\Box+m-1)\chi_{s_0}]_x\cr
&+\alpha^2\sum_{x,y}[\chib_{s_0}(\Box+m-1)]_x(V^\dagger_xP_R+V_xP_L)
\dsl_{xy}(V^\dagger_yP_R+V_yP_L)[(\Box+m-1)\chi_{s_0}]_y\cr
&+O(\alpha^3).
}$$
To order $\alpha$, replacing $V$ by its expectation value $v{\bf 1}$,
the Dirac equation reads (in momentum space)
$$i\ssl(p)\chi_s+a(s)(m-1-F(p))\chi_s+P_R\chi_{s+1}+P_L\chi_{s-1}
-\alpha v(m-1-F(p))^2\delta_{s,s_0}\chi_{s_0}=0.\eqno(diracagain)$$
Effectively, $a(s)$ of \eq{as} no longer vanishes at $s=s_0$.  Instead,
$$a(s_0)=\alpha v(1-m+F(p)),\eqno(asagain)$$
and there are no zeromode solutions for $v\ne 0$.  In the broken phase,
the zeromodes pick up a mass of order $v/y$.

We see that in the broken phase, we effectively obtain a nonzero value
for $a(s_0)$ (\eq{asagain}).  This is only possible in the broken phase.
The symmetry \eq{symmetry} acts on $\chi_{s_0}$ as
$$\chi_{Rs_0}\to g\chi_{Rs_0},\ \ \chi_{Ls_0}\to h\chi_{Ls_0},$$
and this symmetry has to be broken in order to have a nonzero value
for $a(s_0)$.
Thus, a righthanded mode inside the waveguide can mix with a lefthanded
 mode outside the
waveguide only if the symmetry \eq{symmetry} is violated, \ie only in
the broken phase.
Moreover, no righthanded and lefthanded modes which are both inside
the waveguide can mix, because no
lefthanded mode inside the waveguide carries
the same lattice momentum as any righthanded mode inside the waveguide,
and lattice momentum is exactly conserved. (See the last section for further
discussion of this point).

We will illustrate this by showing
that the massless poles which exist at $y\to\infty$ (\cf \eq{propagator})
remain
massless to second order in $\alpha=1/y$, and that therefore the massless
spectrum of the theory remains the same for large but finite $y$.
For $G=U(1)$,
to order $\alpha^2$, the inverse propagator in the symmetric phase
becomes
$$\eqalignno{
S^{-1}_{s,s'}(p)&=S^{(0)-1}_{s,s'}(p)-\Sigma_{s,s'}(p),\cr
\Sigma_{s,s'}(p)&=-i\alpha^2\ssl(p)\delta_{s,s_0}\delta_{s_0,s'}
b^2(p)(L+K(p)),&(exampleprop)
}$$
with
$$L=<V_xV^\dagger_{x+\mu}>\eqno(L)$$
and
$$\int_lD(l-p){{\ssl(l)b^2(l)}\over{s^2(l)+\M^2(l)}}
\equiv\ssl(p)K(p).\eqno(K)$$
$D(p)$ is the scalar propagator.  The $L$ term comes from a diagram with
one order
$\alpha^2$ vertex
(\cf \eq{schi}), and the $K(p)$ term comes from a diagram with two order
$\alpha$ vertices.  Since $\Sigma(p)$ vanishes for all $p=\pi_A$ (\eq{pisubA}),
the massless pole structure of the propagator is not changed by
the $\alpha^2$ correction.
(Note that the function $K$ goes to
a constant for $p\to\pi_A$ in the symmetric phase.)
We believe this result to be valid to all orders in
$\alpha$.

Let us conclude this section by considering the effective action for
$\psi$, \eq{spsi}.  This effective action is exact, and therefore should
describe all zeromodes at the waveguide boundary for all values of the
Yukawa coupling $y$, as long as $\psi$ couples to the field $\chi$, \ie
for $\alpha\ne 0$.  The term linear in $\alpha$ contains the $y\to\infty$
propagator for the field $\chi$, and therefore the zeromodes at the
waveguide boundary at large $y$ are present in the effective action
$S_\psi$.  For small $y$, it was shown in \I\ that there is only a
zeromode at the waveguide boundary for $\pi_A=0$, and this should also
follow from $S_\psi$.  This is indeed the case: by rescaling the field
$\psi$ in a way appropriate for small values of $y$ (\ie undoing the
transformation \eq{rescale}), and choosing $y=0$, we see that the $\psi$
propagator for small $y$ is defined by the order $\alpha$ term in \eq{spsi}
to lowest order in $y$.  In momentum space, this propagator reads
$$
S^{(0)}_\psi(p)={{-i\ssl(p)}\over{s^2(p)}}{{s^2(p)+\M^2(p)}\over
{s^2(p)+\M^2(p)+b^2(p)}}.$$
This expression only has a pole at $p=0$, and none at $p=\pi_A$ with
$\pi_A\ne 0$ (\cf end of section 3), and describes one massless Dirac
fermion at the waveguide boundary, in accordance with the results of
\I\ at small $y$.

\subhead{\bf 6. Conclusion}

In this paper we continued the investigation of the waveguide implementation
of the domain wall fermion model [\cite{kaplan2,paper1}].  In \I\ we
argued that the question as to whether this model can yield a chiral
gauge theory is a dynamical question which cannot be decided on the basis
of simple anomaly arguments.  Therefore, a complete investigation of the
phase diagram is needed in order to see what the massless fermion spectrum
of the model is, as a function of the parameters of the model.  However,
in \I\ this analysis was left incomplete, because of difficulties with the
numerical exploration of the phase diagram for large values of the
Yukawa coupling.

Here we showed that the action for this model can be
reformulated in such a way that a systematic strong coupling expansion in
the Yukawa coupling can be performed.  This allowed us to examine the
strong Yukawa coupling region of the phase diagram analytically.

In the limiting case $\alpha=0$ ($y\to\infty$) the fermions decouple from the
scalar field, and we obtained a free fermion theory which can be solved
exactly. We found a very rich zeromode
spectrum.  While the zeromode spectrum at the domain wall and antidomain
wall is unchanged, in a sense all their doubler modes show up localized
at the waveguide boundaries.  The complete massless fermion
spectrum inside as well as
outside the waveguide is vectorlike.  The gauge fields, which live only
inside the waveguide, will couple to all the massless fermions
inside, and the
result is a vectorlike gauge theory.
The situation is in fact worse than at weak
Yukawa coupling, where inside the waveguide
only one massless mirror mode emerged due to the
interaction of the fermions with the scalar field.

The phase diagram
consists of symmetric phases at weak and strong Yukawa coupling
(for small enough values of the scalar hopping parameter), separated
by a phase in which the gauge symmetry is spontaneously broken.
We verified by an explicit calculation that the massless spectrum
remains unchanged to first order in $\alpha^2$ in the symmetric strong coupling
phase. While we did not provide a rigorous proof, there can be little doubt
that this result extends to all orders in $\alpha^2$.
There is an interesting relation between the present model and
the Smit-Swift model
 [\cite{donreview}].  The massless fermion spectrum at
strong Yukawa coupling is similar to that of the Smit-Swift model
at weak Yukawa coupling, and {\it vice versa}.

  The question arises whether a more general framework could lead to
qualitatively different results. For example, by a suitable modification
of the scalar action, it might be possible to arrange for the existence of
an antiferromagnetic
condensate which is $g$- and $h$-invariant. In that case, lattice momentum
would
be conserved only modulo $\pi$ (and not modulo $2\pi$) and mixing between
righthanded and lefthanded modes from different corners of the Brillouin zone
would be possible inside the waveguide boundary.
Without fine tuning, we would expect that
only a single lefthanded mode would survive inside the boundary. But since
lefthanded and righthanded modes would pair to form massive states, the
full massless spectrum inside the waveguide would remain vectorlike!

  There is a good reason to believe that the phenomenon described in this
example is completely general.
The point is that we start (at $\alpha=0$) with a
well-defined theory that has a relativistic,
albeit vectorlike, low energy spectrum.
The question is whether one can achieve a chiral spectrum by making a
continuous change of parameters which leaves the ($g$ and $h$) symmetries
unbroken. Assuming the model maintains
a consistent continuum limit throughout this process
(which should be the case if we add only finite range operators to the
lattice action),  we expect that the low energy spectrum can be
correctly described by a corresponding change of parameters
in a relativistic effective lagrangian.
But then the only way to decouple massless fermions
is to pair them into massive Dirac fermions. Thus, since we started with a
vectorlike spectrum, we will end up with a vectorlike spectrum.

  We conclude that, although the phase diagram is richer than suspected
in ref.\ \I, the waveguide implementation of the domain
wall fermion model does not lead to a lattice chiral gauge theory.

\subhead{\bf Acknowledgements}

We would like to thank Karl Jansen and
Jeroen Vink for many useful discussions.  M.G.
would like to thank Michael Ogilvie for some discussions.  This work is
supported in part by the Department of Energy under contract
\#DOE-2FG02-91ER40628.

\references

\refis{kaplan}
D.B. Kaplan, \plb{288} (1992) 342.

\refis{narneu}
R. Narayanan and H. Neuberger, \plb{302} (1993) 62.

\refis{shamir}
Y. Shamir, \npb{406} (1993) 90.

\refis{calhar}
C.G. Callan and J.A. Harvey, \npb{250} (1985) 427.

\refis{jansen}
K. Jansen, \plb{288} (1992) 348.

\refis{goljankap}
M.F.L. Golterman, K. Jansen and D.B. Kaplan, \plb{301} (1993) 219.

\refis{crehor}
M. Creutz and I. Horvath, Nucl. Phys. {\bf} (Proc.Suppl.) {\bf 34}
(1994) 583.

\refis{janschm}
K. Jansen and M. Schmaltz, \plb{296} (1992) 374.

\refis{kaplan2}
D.B. Kaplan, Nucl. Phys. {\bf B} (Proc.Suppl.) {\bf 20} (1993) 597.

\refis{paper1}
M.F.L. Golterman, K. Jansen, D.N. Petcher and J.C. Vink, \prd{49} (1994)
1606; Nucl. Phys. {\bf B} (Proc.Suppl.) {\bf 34} (1994) 593.

\refis{narneu2}
R. Narayanan and H. Neuberger, \prl{71} (1993) 3251; \npb{412} (1994)
574; Nucl. Phys. {\bf B} (Proc.Suppl.) {\bf 34} (1994) 587; R. Narayanan,
Nucl. Phys. {\bf B} (Proc.Suppl.) {\bf 34} (1994) 95.

\refis{shamir2}
Y. Shamir, \npb{417} (1994) 167.

\refis{janzak}
J. Smit, Acta Physica Polonica {\bf B17} (1986) 531.

\refis{janseillac}
J. Smit, Nucl. Phys. {\bf B} (Proc.Suppl.) {\bf 4} (1988) 451.

\refis{jannpb}
J. Smit, \npb{175} (1980) 307; L.H. Karsten, in {\it Field Theoretical
Methods in Particle Physics,} ed. W. R\"uhl, Plenum (1980)
(Kaiserslautern 1979).

\refis{swift}
P.D.V. Swift, \plb{145} (1984) 256.

\refis{jan}
L.H. Karsten and J. Smit, \npb{183} (1981) 103.

\refis{donreview}
D.N. Petcher, Nucl. Phys. {\bf B} (Proc.Suppl.) {\bf 30} (1993) 50 and
references therein.

\endreferences

\vfill
\bye